\documentclass[]{aa}
\usepackage{graphicx}
\usepackage{textcomp}
\usepackage{gensymb}
\usepackage[varg]{txfonts}
\usepackage{color}
\usepackage[colorlinks,linkcolor=blue,citecolor=blue,linktocpage=true,
,plainpages=false,urlcolor=blue]{hyperref}

\bibpunct{(}{)}{;}{a}{}{,}
\newcommand{\sunrise}{\textsc{Sunrise}}

\begin{document}
\title{Observations of solar chromospheric heating at sub-arcsec spatial 
resolution}
\titlerunning{}
\authorrunning{Smitha et. al.}
\author{H.~N.~Smitha \inst{1},
L.~P.~Chitta\inst{1}, T.~Wiegelmann\inst{1}, \and  S.~K.~Solanki\inst{1,2}}
\institute{
$^{1}$Max-Planck-Institut f\"ur Sonnensystemforschung, Justus-von-Liebig-Weg 3, 37077 
G\"ottingen, Germany\\
$^{2}$School of Space Research, Kyung Hee University, Yongin, 
Gyeonggi, 446-701, 
Republic 
of Korea\\
\email{smitha@mps.mpg.de}}

\abstract
{A wide variety of phenomena such as gentle but persistent brightening, dynamic slender features ($\sim$100\,km), and compact ($\sim$1\arcsec) ultraviolet (UV) bursts are associated with the heating of the solar chromosphere. High spatio-temporal resolution is required to capture the finer details of the likely magnetic reconnection-driven, rapidly evolving bursts. Such observations are also needed to reveal their similarities to large-scale flares, which are also thought to be reconnection driven, and more generally their role in chromospheric heating. Here we report observations of chromospheric heating in the form of a UV burst obtained with the balloon-borne observatory, \sunrise{}. The observed burst displayed a spatial morphology similar to that of a large-scale solar flare with circular ribbon. While the co-temporal UV observations at 1.5\arcsec\ spatial resolution and 24\,s\ cadence from the Solar Dynamics Observatory showed a compact brightening, the \sunrise{} observations at diffraction-limited spatial resolution of 0.1\arcsec\ at 7\,s\ cadence revealed a dynamic sub-structure of the burst that it is composed of extended ribbon-like features and a rapidly evolving arcade of thin ($\sim$0.1\arcsec wide) magnetic loop-like features, similar to post-flare loops. Such a dynamic sub-structure reveals the small-scale nature of chromospheric heating in these bursts. Furthermore, based on magnetic field extrapolations, this heating event is associated with a complex fan-spine magnetic topology. Our observations strongly hint at a unified picture of magnetic heating in the solar atmosphere from some large-scale flares to small-scale bursts, all being associated with such a magnetic topology.}

\keywords{Sun: atmosphere, Sun: magnetic fields, Sun: photosphere, Sun:chromosphere, magnetic reconnection}
\maketitle

\section{Introduction}
\label{sec:intr}
The solar chromosphere is sandwiched between the cooler photosphere and the multi-million Kelvin corona. High spatial and temporal  resolution observations reveal a richly structured and dynamic chromosphere that is hotter (with plasma temperature in excess of $10^4$\,K in the uppermost layers) than the photosphere. The chromospheric plasma loses energy through radiation ranging from 4\,kW\,m$^{-2}$ in the quiet Sun to 20\,kW\,m$^{-2}$ in active regions; see e.g. \cite{1977ARA&A..15..363W}, which needs to be continually replenished with energy transported from the photosphere.  

The evolution of photospheric magnetic field {governs} the energy transport into higher layers and {plays} an important role in structuring the atmosphere. Consequently, intense heating events in the solar atmosphere, in which the role of magnetic field evolution is readily apparent, show intricate spatial and temporal evolution. For instance, Ellerman bombs in the photosphere \citep{1917ApJ....46..298E} have a flame-like appearance \citep{2011ApJ...736...71W}. These flames are interpreted as plasma jets from the photosphere and have been modelled with radiation magnetohydrodynamic simulations as reconnection events during flux emergence and cancellation \citep{2017A&A...601A.122D,2017ApJ...839...22H}. Often the chromosphere above the sites of flux emergence and cancellation also evolves dynamically. Compact intensity enhancements or bursts (clearly distinguishable in ultraviolet, UV, radiation) and plasma jets are observed at these locations \citep[e.g.][]{2002ApJ...575..506G,2007Sci...318.1591S,2014Sci...346C.315P, 2015ApJ...809...82G, 2015ApJ...812...11V, 2018ApJ...854..174T,2018arXiv180505850Y}. Based on the  extrapolations of photospheric magnetic field, such events are thought to be triggered by the magnetic energy released through reconnection at the base of the chromosphere \citep[e.g.][]{2010ApJ...724.1083G,2017A&A...605A..49C}.

Investigating the internal structures of chromospheric heating in intense UV burst-type events will provide information on the nature of magnetic energy release and its deposition in the lower atmosphere. High resolution imaging and spectroscopic observations enable such a detailed analysis. For example, \citet{2016ApJ...819L...3Z} studied details of magnetic topology and a two-step reconnection process that triggered a small-scale chromospheric jet. \citet{2017A&A...605A..49C} presented a spectroscopic analysis of a UV burst which is composed of spatially decoupled blue and redshifts of emission lines, which they interpreted as plasma flows from the reconnection site, with current sheet inclined to the line of sight. \citet{2017ApJ...851L...6R} observed small-scale 
($<$0.2\arcsec), rapidly evolving (order of seconds) brightenings in chromospheric bursts which they interpreted as evidence for plasmoids that are thought to mediate {the} fast reconnection process. However, the nature of the observed underlying magnetic topology and its association with the internal structure of chromospheric bursts at small spatial scales ($\approx$0.1\arcsec) is not well known. 

Here we present the details of a UV burst in an emerging active region by combining high resolution magnetograms and high cadence chromospheric imaging. Our observations and magnetic field extrapolations show the presence of a small-scale fan-spine magnetic topology, which manifests itself as a bright ribbon-like and transient magnetic loop-like features in the chromosphere. We describe the spatial and temporal properties of chromospheric heating in the burst.

\section{Observations}
\label{sec:Obs}

To investigate the details of chromospheric heating in UV bursts we focus on an emerging active region AR 11768, for which high spatial and temporal resolution observations are available from the balloon-borne \sunrise{} Observatory 
\citep{2010ApJ...723L.127S,2017ApJS..229....2S,2011SoPh..268....1B, 2011SoPh..268..103B,2011SoPh..268...35G,2011SoPh..268...57M}. During its second flight on 2013 June 12 at 23:39\,UT, \sunrise{} observed AR 11768, at a heliocentric angle $\mu=0.93$. The \sunrise{} Filter Imager \citep[SuFI;][]{2011SoPh..268...35G} obtained high cadence ($\approx7$\,s) chromospheric Ca\,{\sc ii}\,H narrow band filter images at 3968\,\AA\ with an image scale of 0.02\arcsec\,pixel$^{-1}$ (diffraction limited resolution of 0.07\arcsec\textendash{} 0.1\arcsec). SuFI observed this AR over 60\,minutes. During this period, SuFI recorded chromospheric brightenings covering an area of about 8\arcsec$\times$ 5\arcsec and lasting for about 20\,minutes. To examine the signatures of these brightenings in the temperature-minimum and the upper photosphere we used time sequences of continuum images at 1700\,\AA\ and 1600\,\AA\ obtained by the Atmospheric Imaging Assembly \citep[AIA;][]{2012SoPh..275...17L,2012SoPh..275...41B} onboard the Solar Dynamics Observatory \citep[SDO;][]{2012SoPh..275....3P}. These data have an image scale of 0.6\arcsec\,pixel$^{-1}$ and a cadence of 24\,s. The AIA 1600\,\AA\ filter also detects emission from the C\,{\sc iv} doublet at 1548\,\AA\ and 1550\,\AA\ formed at around $10^5$\,K.

To study the evolution of the surface magnetic field underlying the brightenings, we used magnetic field maps of AR 11768 retrieved from the  Imaging Magnetograph eXperiment \citep[IMaX;][]{2011SoPh..268...57M} on board \sunrise{}. The IMaX recorded the full Stokes vector of the Fe~{\sc i} 5250.2\,\AA\ line at eight wavelength positions, with a cadence of 36.5\,s\ and an image scale of 0.0545\arcsec\,pixel$^{-1}$. The magnetic field vector is obtained by inverting the IMaX Stokes profiles using the SPINOR \citep{2000A&A...358.1109F}, which makes use of the STOPRO routines \citep{1987PhDT.......251S}. These high resolution magnetic field maps are available only from 23:39\,UT\ to 23:55\,UT\ and did not capture the magnetic field evolution of the chromospheric event during its peak brightness. Thus, we supplement these IMaX data with the line of sight magnetic field maps obtained from the Helioseismic and Magnetic Imager \citep[HMI;][]{2012SoPh..275..207S} onboard SDO. These HMI data cover the full solar disc, have an image scale of 0.5\arcsec\,pixel$^{-1}$ and a cadence of 45\,s. The SDO data have been rotated clockwise by 11.1\degree\ to match the orientation of the \sunrise{} data. 

\begin{figure}
\begin{center}
 \includegraphics[width=0.48\textwidth]{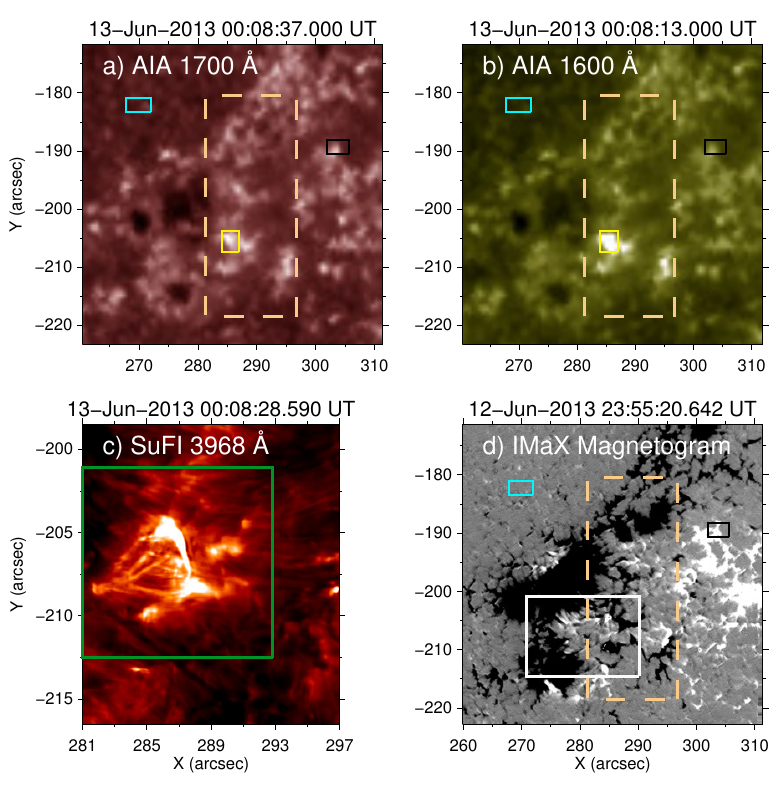}
  \caption{Overview of the chromospheric burst. a) AIA 1700\,\AA\ map showing the upper photosphere, b) AIA 1600\,\AA\ displaying the upper photosphere and traces of transition region plasma at $\sim10^5$\,K due to the contribution from the C {\sc iv} doublet at 1550\,\AA. The solid yellow boxes at ($286\arcsec,-206\arcsec$) in panels a and b mark the location of the burst.  c) SuFI narrow band filter image of Ca\,{\sc ii}\,H at 3968\,\AA\ showing the burst at a factor of 10 higher spatial resolution. A bright ribbon-like feature and several {narrow fibrilar features} can be seen. {Based on the results of the magnetic field extrapolation, we see that these fibrils are associated with small magnetic loops. Therefore, we call them loop-like structures.} The box marks the region displayed in Fig.\,\ref{sufi_tile}. d) Distribution of the line-of-sight component of the magnetic field underlying the burst, obtained by the IMaX and saturated at $\pm250$\,G. The box highlights the region displayed in Fig.\,\ref{mag_tile}. {The solid cyan boxes in panels a,b, and d at ($273\arcsec,-182\arcsec$) cover a quiet Sun region, and the solid black boxes in these panels at ($303\arcsec,-190\arcsec$) cover a magnetic concentration, both outside the SuFI field of view.} This map is the last of the available IMaX observations and hence it is not co-temporal with the other maps. The full field of view of SuFI is indicated by a dashed box in panels a, b, d. Only a part of the SuFI FOV is plotted in panel c. See Sects.\,\ref{sec:Obs} and \ref{sec:morph} for details.}
\label{context}
\end{center}
\end{figure}	


\section{Morphology of the heating event}
\label{sec:morph}

\begin{figure}
\begin{center}
 \includegraphics[width=0.48\textwidth]{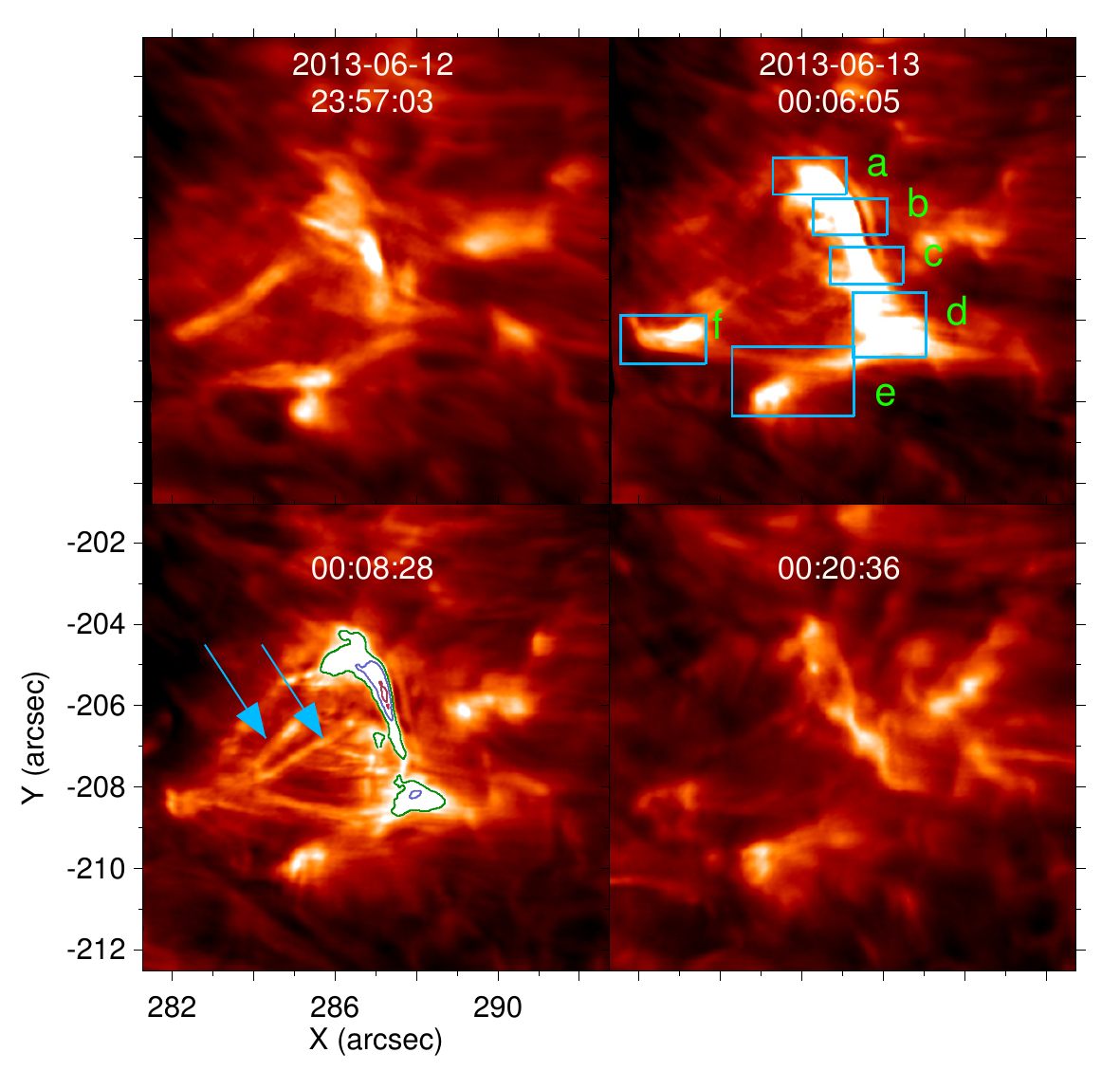}
    \caption{Spatial and temporal evolution of the chromospheric burst. The top left and bottom right panels mark the beginning and end of the heating event. The other two panels show intermediate stages. The boxes a\textendash f are used to compute the average intensities plotted in Fig.\,\ref{lightcurve}. {The outer to inner contours enclose the regions with intensities of $(0.5,0.7,0.9)\times I_{\text{max}}$ in the ribbon-like feature. Here, $I_\text{max}$ is the intensity maximum in the respective SuFI snapshot. The arrows point to loop-like features}, associated with the burst. See Sect.\,\ref{sec:morph} for details. The entire time series of this event is available as an online movie.
\label{sufi_tile}}
\end{center}
\end{figure}

The UV burst under investigation is observed in the vicinity of a group of magnetic pores, which are seen as dark features in the AIA intensity maps tracing the upper photosphere (top panels in Fig.\,\ref{context}). These pores and the surrounding regions constitute predominantly negative polarity magnetic field at the solar surface (lower right panel in Fig.\,\ref{context}). Parasitic positive polarity magnetic field is observed to evolve within a ring of negative magnetic polarity (see white box). A time series of the surface magnetic field maps from HMI (cf. Sect.\,\ref{sec:mtop}, Fig.\,\ref{mag_tile}) shows interaction between the two polarities near ($285$\arcsec,$-205$\arcsec) which is also the location of the UV brightening seen in AIA intensity maps. Here, AIA 1700\,\AA\ and 1600\,\AA\ images display a compact but distinct brightening (see yellow box in Fig.\,\ref{context}). While the brightening recorded in the AIA 1700\,\AA\ channel is comparable to the surrounding bright regions, the intensity of the heating event imaged by AIA 1600\,\AA\ is very prominent compared to its surroundings. However, due to the moderate resolution of the AIA, the internal structure of this heating event could not be investigated further. This event could not be identified in the AIA extreme ultraviolet filter images.

This heating event displays a complex morphology and internal structure when observed near the diffraction limit of the 1-m diameter \sunrise{} telescope with the narrow-band filter of SuFI at 3968\,\AA\ (lower left panel in Fig.\,\ref{context}). In SuFI observations, the event starts at 12 June 2013 23:55\,UT\ and lasts till the end of the time series, with a brief fading around 00:18\,UT\ on 13 June 2013. It reaches maximum brightness on 13 June 2013 around 00:06\,UT. Two distinct features associated with the heating event can be identified in this map. First, an extended bright structure (ribbon-like feature) near the center of the field of view, and second, thin strands or {fibrilar features} that connect the extended bright structure with a point-like region near ($282$\arcsec,$-208$\arcsec). Extrapolation of the magnetic field (cf. Sect.\,\ref{sec:mtop}) at the site of the burst reveals several closed magnetic field lines down to 100\,km\textendash 200\,km above the surface. These closed field lines are found in the same restricted area as the observed fibrilar features in SuFI. Therefore, we refer to these fibrils as loop-like features as seen from above. Based on the magnetic complexity in that region (see Fig.~\ref{context}d), however, we emphasize that it is not possible to show here a one-to-one correspondence between the traced magnetic field lines from extrapolations and the observed loop-like features.


The high spatial resolution of SuFI allowed us to study the temporal evolution of the spatial structure of the heating event in detail (Fig.\,\ref{sufi_tile}, see also the online movie). Prior to the burst, the extended bright region and magnetic loop{-like features} that are bright (i.e. magnetic loops loaded with plasma) are not present (top left panel). Rapid brightening of the region is seen over the course of the next 10 minutes {(with intensity peaking at 2013-06-13 00:06\,UT; cf. Sect.\,\ref{sec:lcurve})}, with the formation of an extended  ribbon{-like feature} (a\textendash d) and compact bright features (e and f). {This ribbon-like feature (outlined by a set of contours on the snapshot at 2013-06-13 00:08\,UT) exhibits a finer substructure with an intensity core near ($287\arcsec,-206\arcsec$). Similar core at the same location is also observed during the time of the intensity peak\footnote{The intensity in SuFI snapshots is scaled for a better visibility of all the features in the figures including the movie. For this reason, the ribbon-like structure does appear saturated and featureless.}. The whole ribbon lies directly on top of a negative polarity magnetic field patch, hinting that its morphology is governed by the underlying magnetic structure (cf. Fig.\,\ref{mag_tile}. See Appendix\,\ref{app:a1} for further discussion).} In the next two minutes, the thin magnetic loop{-like features} appear transiently from the extended region to one of the compact regions (f). At the end of the heating event, the region returns to the state prior to the event, i.e. devoid of bright features and thin loops (lower right panel in Fig.\,\ref{sufi_tile}). 

Some of the loop-like features are seen to extend to nearby bright structures. These chromospheric brightenings are related to the emergence of magnetic flux observed at two sites adjacent to the heating event, one near the edge of the SuFI FOV and the other outside, both on its right side \citep{2017ApJS..229....3C}. In between the burst studied here and one of the flux emergence events, an Ellerman bomb is triggered which reaches its maximum brightness around 12 June 2013 23:47\,UT \citep{2017ApJS..229....5D}. Thus the region captured by SuFI is highly dynamic with bright loop-like features \citep[or slender fibrils,][]{2017ApJS..229....6G} stretching across the FOV from one region to the other. Upon visual inspection, it appears that the brightening of the heating event follows the brightening of the fibrils, connecting this event to the flux emergence site on the right (see online movie). However, the complex magnetic field evolution observed in the HMI and IMaX magnetograms reveal the possible origin of this burst due to local interactions of positive and negative magnetic polarities at the site of the burst. We will discuss the origin of this burst and its possible link to the magnetic topology in Sect.\,\ref{sec:mtop}.


\section{Light curves from the heating event}
\label{sec:lcurve}

\begin{figure}
\begin{center}
 \includegraphics[width=0.48\textwidth]{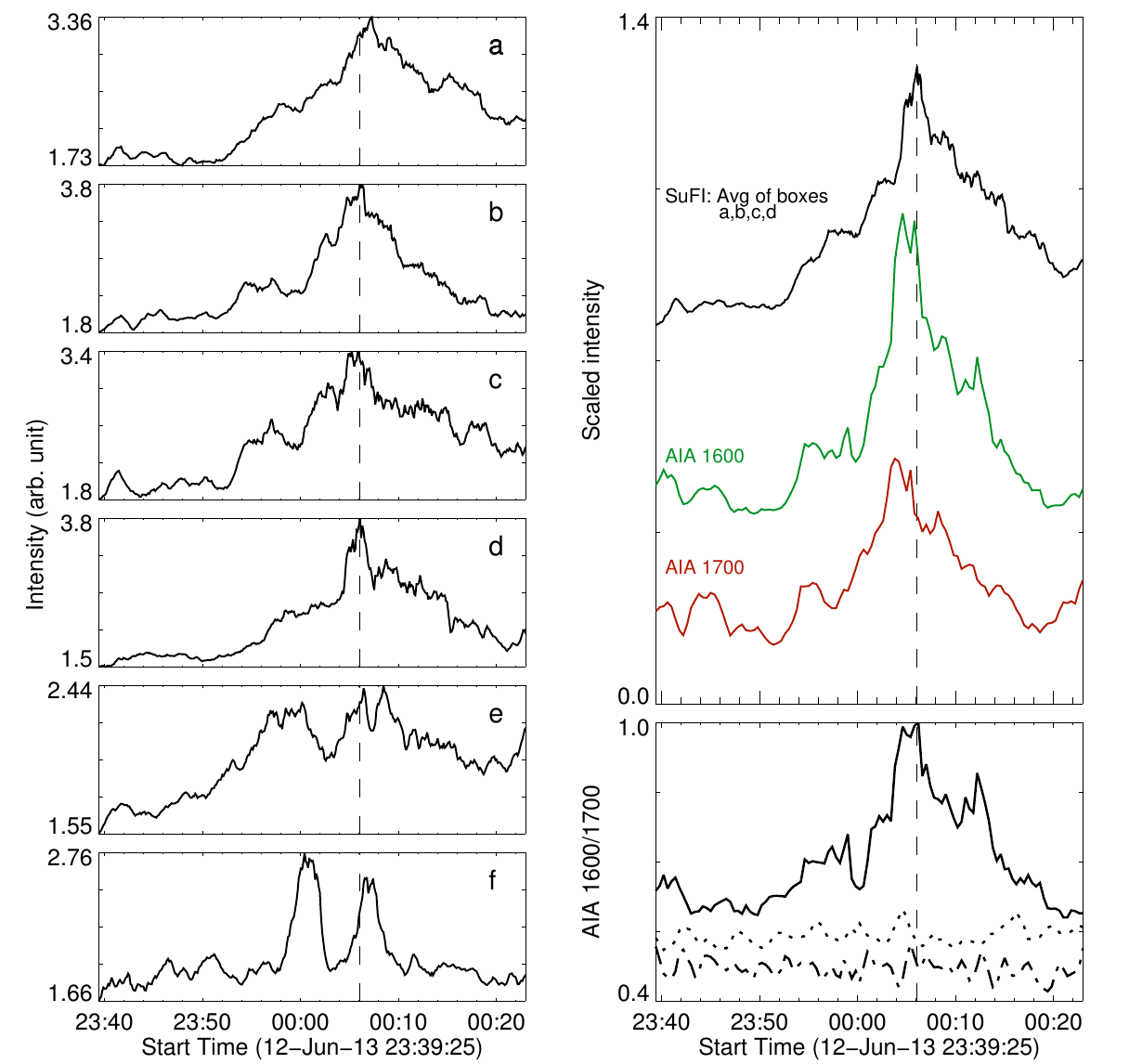}
 \caption{History of the burst from different sub-regions. Left: SuFI light curves  a\textendash f are average intensities from different sub-regions of the burst (see boxes marked a\textendash f in Fig.\,\ref{sufi_tile}).  Top right: Plotted in black is the average SuFI intensity from boxes marked a\textendash d in Fig.\,\ref{sufi_tile}. Green and orange light curves are average intensities as recorded by AIA 1600\,\AA\ and AIA 1700\,\AA\ filters from the solid yellow box, covering the burst displayed in Fig.\,\ref{context}. The {solid curve in the} bottom right panel is the ratio of average {burst} intensities from AIA 1600\,\AA\ and AIA 1700\,\AA\ filters, normalized to unity {at the maximum}. {The dash-dot curve is the same ratio but for the filter ratio of average quiet Sun intensities from the solid cyan box in Fig.\,\ref{context}, and the dotted curve is for the magnetic concentration indicated by the black box. These two ratio curves are normalized to the maximum of the burst ratio curve to highlight the relative differences in the intensity ratios from the three regions.} Vertical dashed lines are plotted as a guide to show the temporal correspondence between various light curves. See Sect.\,\ref{sec:lcurve} for details.}
\label{lightcurve}
\end{center}
\end{figure}

The morphology of the chromospheric UV burst discussed in Sect.\,\ref{sec:morph} suggests a possible temporal correlation of the energy release over an extended region. To investigate this temporal correspondence between different regions in the burst, we have spatially divided it into six sub-regions (see Fig.\,\ref{sufi_tile}). 

Spatially averaged intensity from each sub-region as a function of time is plotted in the left panel of Fig.\,\ref{lightcurve}. Light curves from different sub-regions within the extended ribbon (b\textendash d) show near simultaneous peaks, while the sub-region (a) peaks a minute later. Light curves from the compact bright features (e and f) show multiple peaks, some of which are near simultaneous with the peaks from the extended region. 

Note that AIA 1700\,\AA\ and 1600\,\AA\ filter images displayed co-spatial brightenings (cf. Fig.\,\ref{context}). To compare intensity variations observed in SuFI data and the AIA data, first we averaged the intensity from sub-regions (a\textendash d), which is plotted as a function of time in the top right panel of Fig.\,\ref{lightcurve}. The light curve shows a single well-defined peak around 00:06\,UT. Both the AIA light curves also show intensity enhancements from the heating event, revealing its role in disturbing the photospheric plasma (similar to Ellerman bombs). 

{These AIA intensity enhancements could be purely due to the enhanced UV continuum emission from the burst. Although these filters sample slightly different heights near the photosphere and may respond differently to temperature gradients, the spectral lines forming at transition region temperatures in the UV will also be highly enhanced in these heating events \citep{1994ApJ...431L..55B,2014Sci...346C.315P}. In fact, the AIA 1600\,\AA\ filter includes the strong C\,{\sc iv} doublet that forms at $10^5$\,K. Moreover, the effective area of that filter has its peak close to the rest wavelengths of C\,{\sc iv} doublet around 1550\,\AA. Thus we expect that the burst here could be heated to even $10^5$\,K. To check for this possibility, we adopt the filter ratio of AIA 1600\,\AA\ to 1700\,\AA\ as a proxy of the C\,{\sc iv} emission. Indeed, we observe a clear peak in the filter ratio curve, again, near simultaneous with the SuFI data (solid curve in the lower right panel). To place this filter ratio curve from the burst in context, we compare it with the same proxy from two different regions (lower right panel). (1) from a quiet Sun region (dash-dot curve) and (2) from a magnetic concentration (dot curve). Due to the little or weak contribution from the C\,{\sc iv} doublet in the quiet Sun, the filter ratio shows on average a constant profile with time. Though at the magnetic concentration the ratio shows some enhancement over the quiet Sun, it remains constant over time. This is because the magnetic concentrations have enhanced visibility in the UV continua \citep{2016A&A...592A.100R}. Therefore, although we cannot rule out a time-dependent change of the photospheric temperature gradient at the location of the burst, the additional enhancement in the filter ratio we observe at the burst could well be due to the $10^5$\,K\ plasma.\footnote{A forward modeling of the UV radiation is required to properly disentangle the C\,{\sc iv} contribution from the UV continuum enhancement. But this is beyond the scope of the present work.}} Thus the chromospheric brightening observed with the SuFI combined with intensity enhancements in AIA 1600\,\AA\ and 1700\,\AA\ and their filter ratio highlight the multi-thermal nature of the heating event.

The near-simultaneity of the brightenings observed from different sub-regions (both extended and compact) of the burst suggests a complex magnetic connectivity between these various sub-regions. Such a temporal correspondence is also observed as sequential brightening seen in circular ribbons of large-scale flares \citep[e.g.][]{2009ApJ...700..559M, 2012ApJ...760..101W}. In these large flares, the sequential brightening is governed by a dome-like magnetic field topology above the surface, which is also the case for the burst here (cf. Sect.\,\ref{sec:mtop}).


\section{Thin magnetic loops}
\label{sec:loops}

\begin{figure*}
\begin{center}
\includegraphics[width=0.78\textwidth]{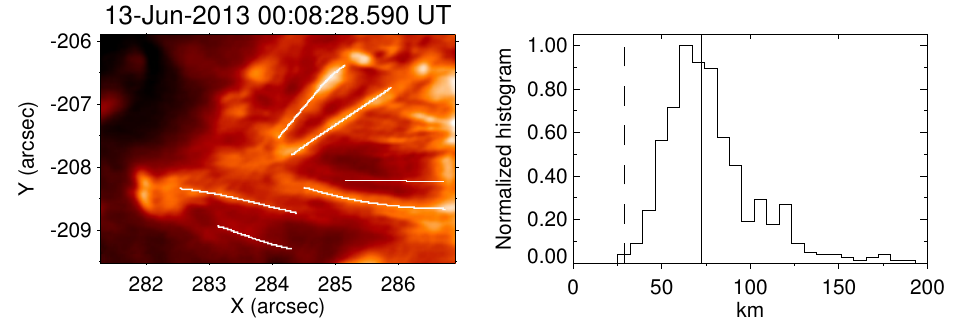}
\caption{Magnetic loop-like features associated with the burst studied here. 
Left:  A sub-field-of-view of SuFI map from Fig.\,\ref{sufi_tile} showing the 
loops. A few of the traced structures are highlighted with solid lines. Right: 
Distribution of widths of selected loops along their length. The distribution 
shows the full width at half maximum of Gaussian fits to the intensity profile 
across a given loop, at each position along the loop. The dashed and solid 
vertical lines {respectively} denote twice the pixel scale and diffraction 
limited spatial resolution of SuFI at 3968\,\AA. See Sect.\,\ref{sec:loops} for 
details.
\label{pfloop}}
\end{center}
\end{figure*}

During the period when the burst reached its peak intensity, we observed transient brightening of several magnetic strands with apparent loop-like configuration as seen from above (left panel in Fig.\,\ref{pfloop}; cf. Fig.\,\ref{sufi_tile}). These resemble the large scale loops seen typically in coronal observations, but at much smaller spatial scales. Such miniature loops (lengths of $\sim$1\,Mm) are also detected in High-resolution Coronal Imager (Hi-C) observations, albeit at coronal temperatures \citep{2013A&A...556A.104P, 2017A&A...599A.137B}. The SuFI magnetic loop-like features observed here could not be identified in coronal images of AIA, indicating that these form below coronal temperatures. These loop-like structures are present only for two minutes.

A striking feature of these magnetic loops is their clear  visibility against chromospheric background emission, which highlights their apparent narrow width. To determine the width of these structures, first we use the automated pattern recognition code by \cite{2010SoPh..262..399A}, tuned to identify loops with a minimum length and curvature radius of 25 and 100 SuFI pixels, respectively. The code detected several curvilinear loop-like features. A sample of these are shown in the left panel of Fig.\,\ref{pfloop}. All the curvilinear loop-like features are then straightened by interpolating intensities onto a rectilinear grid. We assume that the intensity profile across each loop segment (along its length) is a Gaussian, \citep[see e.g.][]{2016ApJ...826L..18B}. Next we compute the loop width, which we define as the full width at half maximum (FWHM) of the best-fit Gaussian intensity profile with a background term, at each loop segment along its length. A histogram of the computed FWHMs is shown in the right panel of Fig.\,\ref{pfloop}. The histogram sharply peaks around 60\,km\textendash 70\,km and drops rapidly towards larger widths. Note that the measured widths peak close to the diffraction limited resolution, which suggests that in actuality at least some of these loops are narrower and remain unresolved in present observations. {For comparison, the width of extended ribbon-like feature is in general a factor of 2--4 larger than that of the loops (see Appendix\,\ref{app:a1}).}

The widths of chromospheric loop-like features reported here are comparable to the widths of bright chromospheric fibrils, in the range of 80\,km--100\,km, studied by \citet{2009A&A...502..647P}, but are smaller than the widths of many other elongated  structures in the solar atmosphere. For instance, using the same SuFI data, \cite{2017ApJS..229....6G} measured the average width of the slender Ca\,{\sc ii} fibrils to be 180\,km\ with the smallest width measured being 100\,km. \citet{2012ApJ...745..152A} measured an average FWHM of 310\,km of the cool plasma condensations formed in coronal rain. The low-lying transition region loops observed by the Interface Region Imaging Spectrograph, at a spatial resolution of 0.32\arcsec, have a FWHM of about 300\,km \citep{2016ApJ...826L..18B}. Using AIA and Hi-C coronal images, \citet{2017ApJ...840....4A} reported that equivalent widths of coronal loops peak at around 550\,km. 

The width of magnetic loops can provide hints on the nature of the heating processes involved in energizing those structures. Based on the width of coronal loops, which is comparable to photospheric granulation, \citet{2017ApJ...840....4A} argued that the coronal loops are energized by heating on macroscopic scales (i.e. comparable to the granular scale) rather than on much smaller scales that are currently unresolved (nanoflares). Although chromospheric loop-like features presented here are quite different from their coronal counterparts, their unresolved nature suggests that the heat deposition, particularly in low-lying bursts \citep[cf. Sect.\,\ref{sec:mtop}; see also][]{2017A&A...605A..49C}, may occur on much smaller spatial scales (i.e. on scales of tens of km as compared to granular scales of hundreds of km).


\section{Magnetic topology}
\label{sec:mtop}

\begin{figure*}
\begin{center}
\includegraphics[width=0.48\textwidth]{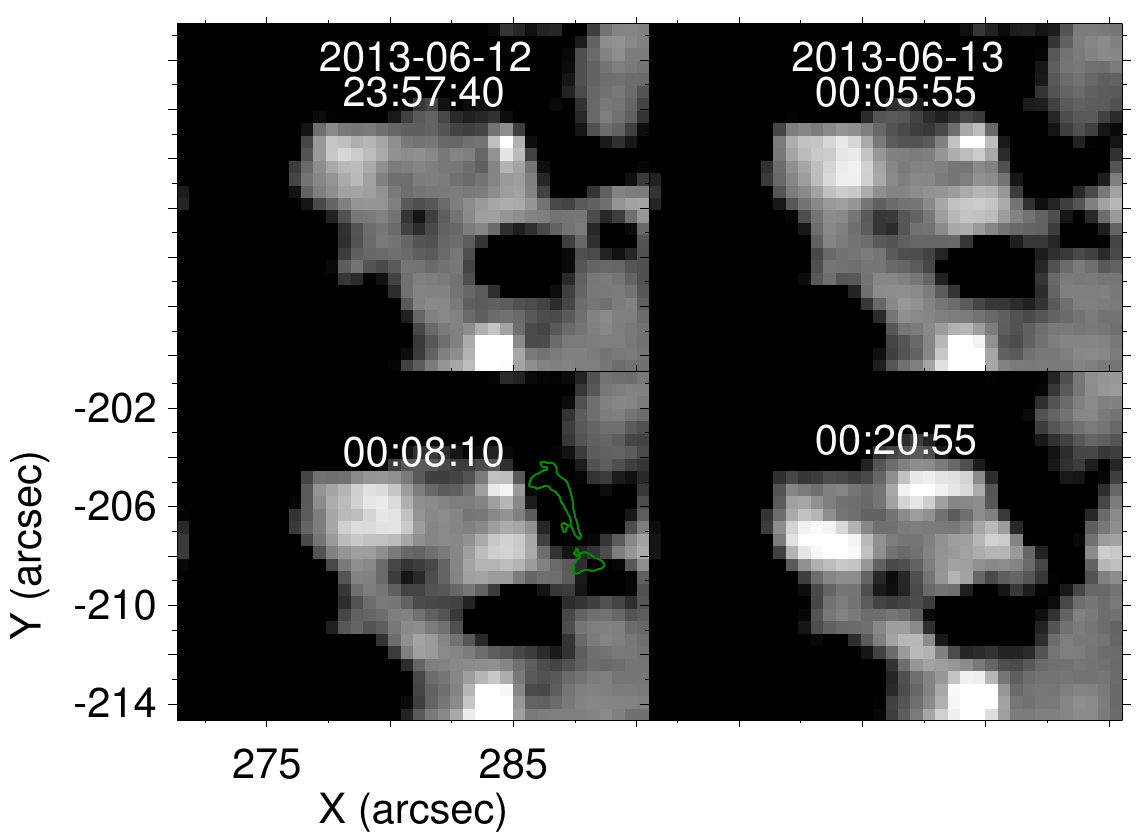}
\includegraphics[width=0.48\textwidth]{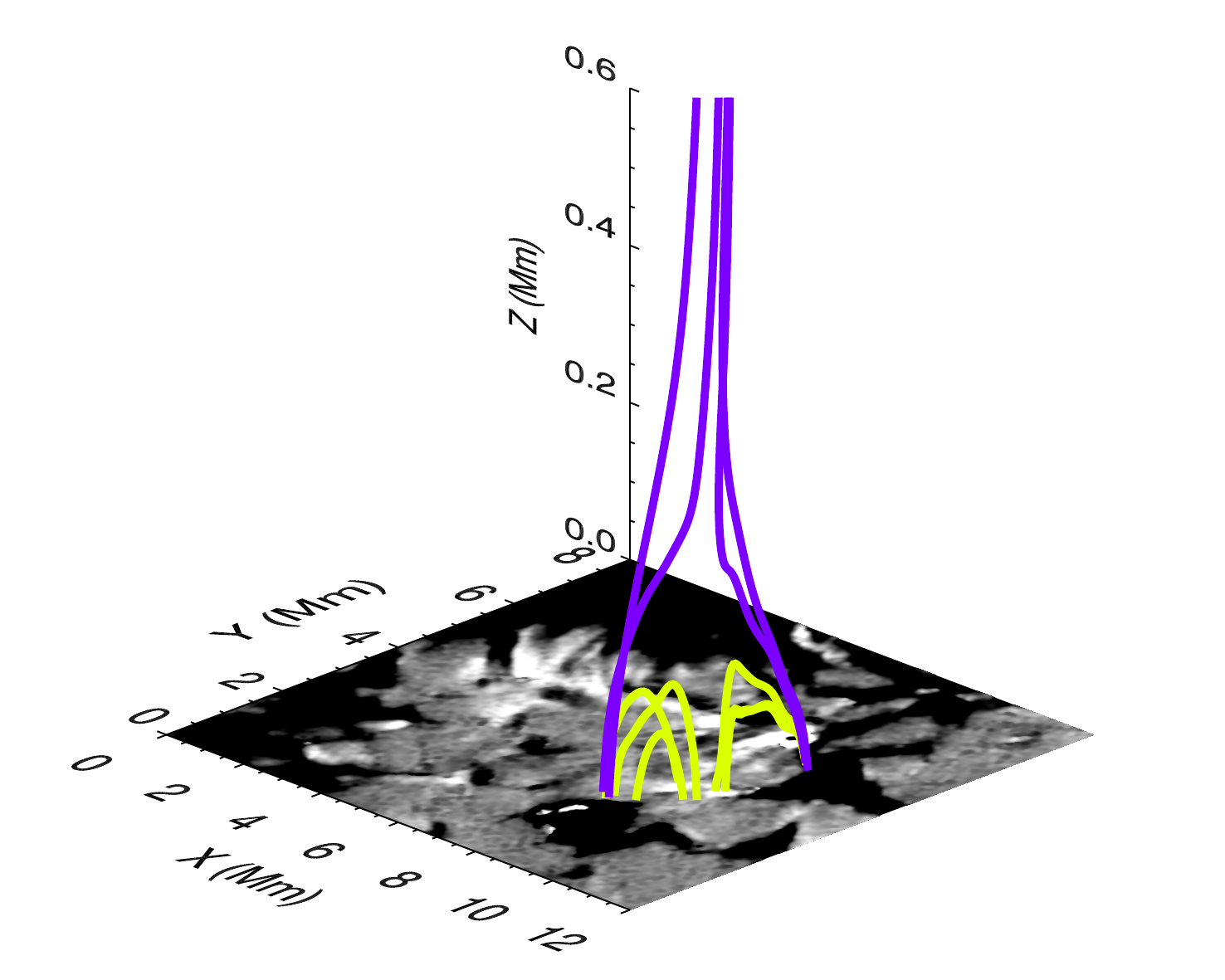}
    \caption{Distribution of the surface magnetic field underlying  the burst. Left: A sequence of HMI line of sight magnetic field maps, near-simultaneous to the SuFI images in Fig.\,\ref{sufi_tile}, saturated at $\pm100$\,G. The FOV here is marked by the solid white box in the IMaX map in Fig.\,\ref{context}. {The contour is the same as the outer contour in Fig.\,\ref{sufi_tile} that outlines the ribbon-like feature observed in SuFI.} Right: Grey scale image is a sub field of view of the IMaX LOS magnetic field (boxed region from Fig.\,\ref{context}). Selected magnetic field extrapolations are plotted in blue (open/spine) and yellow (closed/dome) to show the magnetic topology surrounding the burst. See Sect.\,\ref{sec:mtop} for details.
\label{mag_tile}}
\end{center}
\end{figure*}

As discussed in Sect.\,\ref{sec:morph}, visual inspection of the HMI (Fig.\,\ref{mag_tile}) and IMaX (Fig.\,\ref{context}) magnetograms suggests that the UV burst is triggered during the interactions between positive and negative polarity magnetic fields. In this case, the negative polarity magnetic region formed a ring-like feature and the positive polarity magnetic elements are embedded within this ring, resulting in a complex magnetic topology at the surface. Clearly, some form of magnetic reconnection is expected in this situation and the energy released during this reconnection process is responsible for the observed chromospheric brightening. 

We have used extrapolations of the surface magnetic field from IMaX  to examine the magnetic topology above the surface that is responsible for this burst. Unlike the nonlinear force-free field extrapolations in which the plasma pressure and gravity are neglected and the magnetic field is assumed to be in a force-free state (i.e. zero plasma-$\beta$), the extrapolations employed here are based on magnetostatic modeling in which these non-magnetic forces are not neglected. {While the force-free assumption is valid in the corona above active regions, this is not the case in the photosphere and chromosphere. In these lower layers of the solar atmosphere, regions with low and high plasma-$\beta$ are present close to each other on small spatial scales. For a meaningful modelling, the non-magnetic forces have to be considered and by invoking} a non-zero plasma-$\beta$ (i.e. the ratio of plasma pressure to magnetic pressure), magnetostatic modeling better represents the solar atmosphere closer to the 
solar surface and thus the magnetic topology governing the bursts closer to the surface. The details of the magnetostatic modeling are described in 
\citet{2015ApJ...815...10W,2017ApJS..229...18W}.

The low resolution HMI magnetograms show that during the progression  of the burst, the overall distribution of the surface magnetic field has not changed drastically (left panel in Fig.\,\ref{mag_tile}), indicating the presence of a stable magnetic topology for several minutes at the surface (although, this surface stability does not necessarily exclude eruptive phenomena higher up in the atmosphere). Therefore, to examine the magnetic topology surrounding the burst, we made use of the last snapshot from the high resolution IMaX data. This snapshot is recorded 13 minutes before the burst reached its peak intensity. 

The extrapolations reveal the presence of a three-dimensional fan-spine magnetic topology \citep{1996PhPl....3..759P} at the site of the burst, including a null point above the surface (right panel in Fig.\,\ref{mag_tile}). A sample of closed (yellow) and open (blue) field lines are drawn to highlight the domains of closed fan surface (dome) and the open spine. The closed field lines connect the embedded parasitic positive polarity magnetic features to the surrounding negative polarity magnetic structure. Theoretical studies and numerical models \citep[see review by][]{2011AdSpR..47.1508P} and field extrapolations based on observed magnetic field \citep[e.g.][]{2009ApJ...700..559M, 2012ApJ...760..101W,2010ApJ...724.1083G,2017A&A...605A..49C} suggest that such a fan-spine magnetic configuration can trigger magnetic reconnection. Here, too, the burst is likely initiated by the reconnection at the null point due to the emergence and cancellation of magnetic flux enclosed by the fan surface. A similar topology is also invoked to explain some long-duration UV bursts which are magnetically coupled to the corona \citep[][also see \citealt{2017ApJS..229....4C}]{2017A&A...605A..49C}. 

{Given that the various parts of the burst are all connected by the same magnetic topology, the apparent difference in its two structures (i.e. the ribbon and the loops) is interesting to note: the ribbon-like feature exhibits a persistent brightening over an extended period of time (more than 10\,minutes) compared to the shorter timescales of the loop-like features (about 2\,minutes for individual features). One reason for this difference could be that reconnection takes place very locally at the null point, so that only individual loops are heated, which then rapidly cool down and become indistinguishable from the background. The persistent brightening of the ribbon is then due to other loops getting heated, as reconnection takes place at different places along the null point.} 

\section{Conclusion}
\label{sec:conc}
In this paper, we have described the fine details of chromospheric heating in a UV burst using diffraction limited ($\approx$0.1\arcsec) observations recorded by \sunrise{}/SuFI. {The burst displayed chromospheric sub-structure with an extended ribbon-like feature with a core and transient magnetic loop-like structures. These intensity features are associated with a fan-spine magnetic topology that is apparent from the magnetic field extrapolations using \sunrise{}/IMaX data. Overall, these intensity (ribbons and transient loops), and magnetic (fan-spine topology) characteristics of the burst resemble some large solar flares with circular ribbon and post-flare loops as seen from above. Thus, our observations point to a unified scenario of complex magnetic coupling through the solar atmosphere, and its role in atmospheric structuring and heating in some large-scale and small-scale bursts.}

Analysis of the evolution of the internal structure of the burst provided clues on the nature of the magnetic energy release. First, the temporal correspondence in terms of intensity variations from different sub-regions of the burst suggest that they are all connected by a complex magnetic topology. This is supported by our finding that the burst is triggered in a fan-spine magnetic topology, low in the solar atmosphere. Second, the transient brightening of magnetic loops, which have widths very close to the diffraction limit of 70\,km\ suggests that the energy deposition in these loops is mainly at small spatial scales. {In addition, we find that the extended ribbon is persistent and wider\footnote{Note that here we cannot rule out the possibility that the ribbon is composed of numerous unresolved loop-like structures (see Appendix\,\ref{app:a1} for further discussion).} compared to the loops.} This raises the question: Which aspects of the chromosphere set the spatial scale of heat deposition in these bursts? Furthermore, a recent study using high resolution observations of an emerging flux region reported that the majority of chromospheric radiative losses in that region are associated with more gentle and persistent heating rather than intense bursts \citep{2018A&A...612A..28L}. Then, what is the magnetic nature of this persistent chromospheric heating in active regions? Does it also constitute small-scale events of energy release in tiny unresolved loops similar to those observed in bursts? These questions remain to be answered.

\begin{acknowledgements}
{We thank the anonymous reviewer for constructive comments that improved the paper.} We acknowledge discussions with R. Gafeira on the SuFI dataset, and S. Jafarzadeh for the derotated SuFI data. H.N.S. acknowledges the financial support from the Alexander von Humboldt foundation. L.P.C. received funding from the European Union's Horizon 2020 research and innovation programme under the Marie Sk\l{}odowska-Curie grant agreement No.\,707837. TW acknowledges DFG-grant WI 3211/4-1. This project has received funding from the European Research Council (ERC) under the European Union's Horizon 2020 research and innovation programme (grant agreement No. 695075), and is supported by the BK21 plus program through the National Research Foundation (NRF) funded by the Ministry of Education of Korea. The German contribution to \sunrise{} and its reflight was funded by the Max Planck Foundation, the Strategic Innovations Fund of the President of the Max Planck Society (MPG), DLR, and private donations by supporting members of the Max Planck Society, which is gratefully acknowledged. The Spanish contribution was funded by the Ministerio de Econom\'ia y Competitividad under Projects ESP2013-47349-C6 and ESP2014-56169-C6, partially using European FEDER funds. The HAO contribution was partly funded through NASA grant number NNX13AE95G. SDO data are the courtesy of NASA/SDO and the AIA, and HMI science teams. This research has made use of NASA's Astrophysics Data System.
\end{acknowledgements}


 \begin{appendix} 
 \section{Width of the ribbon}
 \label{app:a1}

\begin{figure*}[htbp]
\begin{center}
\includegraphics[width=\textwidth]{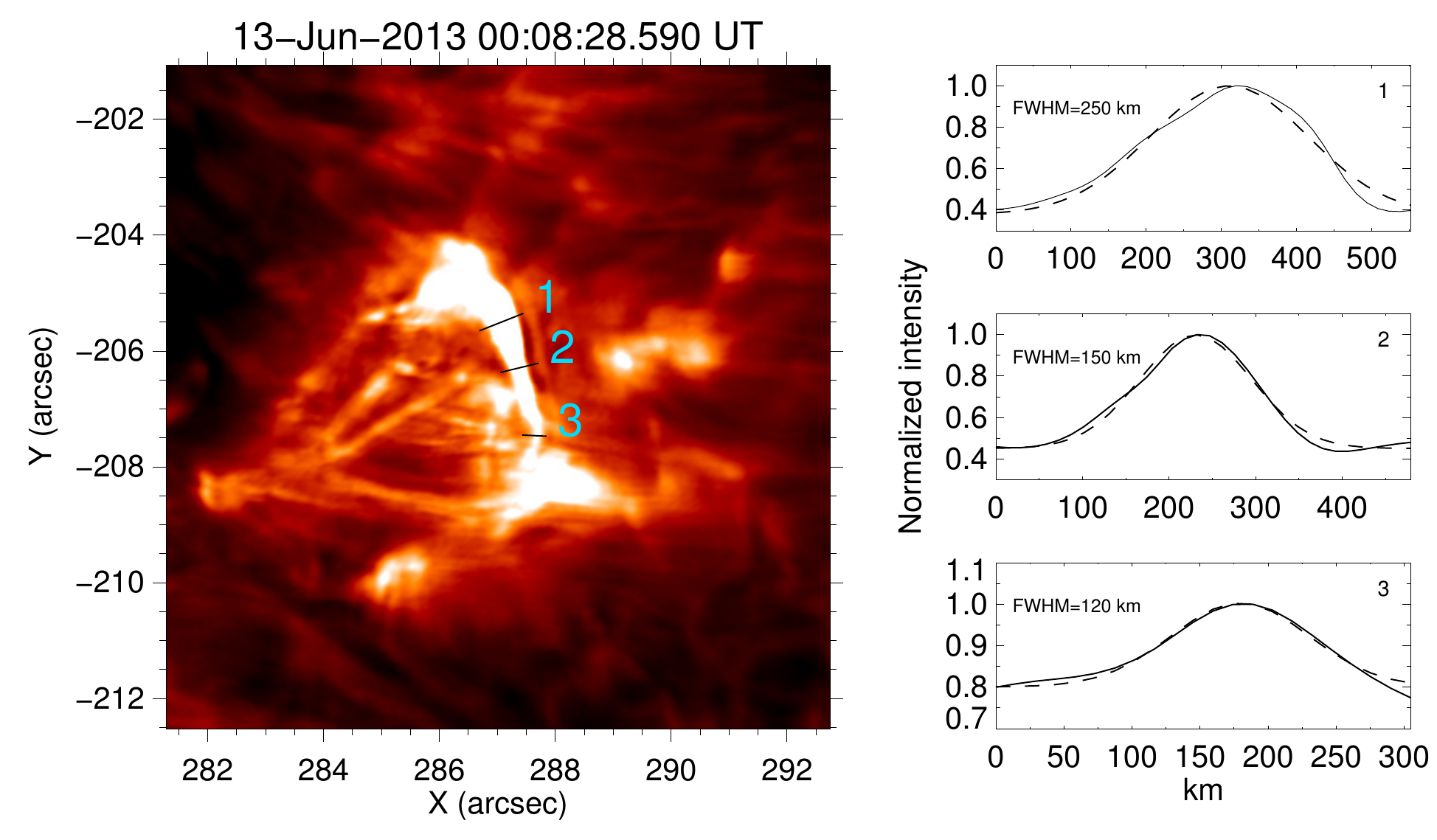}
\caption{{Extended ribbon-like feature associated with the burst. The left panel covers the full extent of the burst and is at the same time-step as the left panel in Fig.\,\ref{pfloop}. Three sample ribbon segments are marked by black lines, numbered 1--3. The intensity profiles (solid) and their Gaussian fits (dashed) along the three ribbon segments are shown in the right panels. The FWHMs of  the three ribbon segments are quoted in the respective panels. See Sect.\,\ref{sec:loops} for details.}
\label{ribbon}}
\end{center}
\end{figure*}
 
{In Sect.\,\ref{sec:loops} we show that the magnetic loop-like features associated with the burst have a distribution of widths peaking close to the diffraction limit of 70\,km. Note that the burst also comprises a ribbon-like feature with a core (see Fig.\,\ref{sufi_tile}). To compare the width of this ribbon with that of the loops, we consider the same time-step that we used to measure the loop widths and compute the width of the ribbon at various positions (as the full width at half maximum of ribbon segments). The width of three sample segments (numbered 1--3) are shown in Fig.\,\ref{ribbon}, which show a FWHM of 250\,km near the top, 120\,km in the narrow part of the ribbon and 150\,km in between (the segment which is closer to the intensity core of the ribbon). We find that the ribbon width varies along its length from 120 km to 240 km, i.e. by a factor of 2 to 4 compared to that of the loops (which is 60 km to 70 km as revealed by the peak of the loop width distribution). In some sections, particularly at both ends, the ribbon is clearly much wider. To provide a possible explanation for the width of the ribbon, we over-plot the intensity contour of the ribbon on the co-temporal HMI magnetogram (Fig.\,\ref{mag_tile}), which shows that the ribbon lies directly on top of a negative polarity magnetic field patch. The HMI observations do not show any obvious concentration of the magnetic field at the location of the SuFI ribbon core. It is possible that there are unresolved magnetic structures invisible to the HMI at this location \citep[see][for examples of such unresolved opposite polarity fields at another location in the SuFI FOV]{2017ApJS..229....4C}. Therefore, we suggest that the appearance of this ribbon and its core are governed by the underlying magnetic structure. In addition, the HMI magnetic field maps do show that there are intermittent smaller positive polarity patches interacting with the negative polarity patch (Fig.\,\ref{mag_tile}). So, we cannot rule out the possibility that in part, the ribbon is composed of numerous unresolved loop like structures.}

 \end{appendix}

\end{document}